# Geometry reconfiguration of a reflector with an auxetic material surface

Bin Xu[1]
*China Academy of Space Technology (Xi'an), Xi'an, China 710000*

Houfei Fang[2], Shuidong Jiang[3], Yangqing Hou[4], Lan Lan[5]
*Shanghai YS Information Technology Co., Ltd., Shanghai, China，201100*

**Mechanically reconfigurable reflectors can modify their geometries to realize large angle beam scanning. This paper studies a method for reconfiguring a reflector geometry from a standard parabolic shape to offset parabolic shapes with varies offset angles. To fulfil the geometry control, the coordinate change of the nodes on the reflector surface are studied with the coordinate transformation method. To accommodate the large deformation of the reflector surface, a negative Poisson's ratio (NPR) material is applied to the reconfigurable reflector.**

## Nomenclature

| | | |
|---|---|---|
| $f$ | = | Focal length of the paraboloid reflector |
| $D$ | = | Aperture diameter of the standard paraboloid reflector |
| $D_\varphi$ | = | Dimeter of the cutting cylindrical surface for offset paraboloid |
| $\varphi$ | = | Offset angle of the offset paraboloid |
| $x_\varphi$ | = | $x$ coordinate of the central axis of cutting cylindrical surface for offset paraboloid |
| $E_{mat}$ | = | Young's modulus of material |
| $E_{eff}$ | = | Effective Young's modulus of material |
| $v_{eff}$ | = | Effective Poisson's ratio |
| $\sigma_x$ | = | Stress in the $x$ dimension |
| $\varepsilon_x, \varepsilon_y$ | = | Strain in the $x$ dimension and $y$ dimension |
| $F$ | = | Force |
| $a,b,g$ | = | Dimensions of an elliptic void |
| $t$ | = | Thickness |
| $\Delta x, \Delta y$ | = | Dimensions changes in the $x$ dimension and $y$ dimension |

## I. Methods of the reflector reconfiguration

As the reflector is axisymmetric, only the scanning in the two-dimensional coordinate *xoz* is discussed in this paper.

The scanning reflector can receive signals from vary angles within certain limits by reshaping the reflector surface. Its working principle is shown in Fig.1. The radio wave parallel to the central axis of the paraboloid focuses on the focal point of the reflector surface, as shown in Fig.1(a). When the radio wave direction is unparallel to the central axis, the reflector should be rotated to the radio wave direction or the focal point cannot receive the signal, as shown in Fig.1(b).

---

[1] Dr.-Ing, xubin84115@163.com.
[2] Principal Engineer, Department of research and development, houfei_fang@yahoo.com, AIAA Senior Member, Corresponding Author.
[3] Dr.-Ing, Department of research and development.
[4] Dr.-Ing, Department of research and development.
[5] Dr.-Ing, Department of research and development.

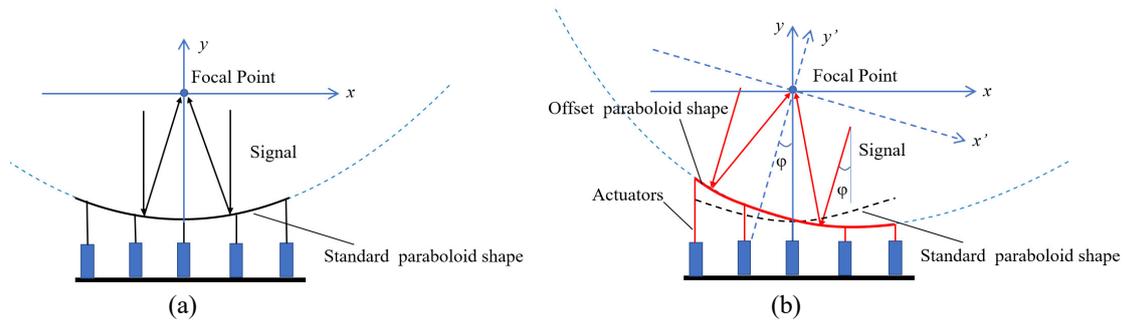

(a)    (b)
Fig.1 The principle of MMR for scanning antenna

For a mechanically reconfigurable reflector, the geometry can be changed between original and offset paraboloid shapes by the actuators. If the center line of the offset paraboloid shape is rotated with respect to the focal point to the original orientation of the standard paraboloid, there is no need to change the position of the antenna feed anymore. As a result, it can save the displacement control system for the feed.

The offset paraboloid geometry can be achieved by cutting the original paraboloid with a cylindrical cutting surface, as shown in Fig.2(a). The diameter of the cylindrical cutting surface is identical to that of the standard paraboloid shape. However, this approach has two drawbacks. The first one is that the deformation from Standard paraboloid shape to offset paraboloid shape is large, as shown in Fig.2(b), that requires the actuators to have large stroke and force. The second one is that the fixed rim require, which is very important for large reflector, cannot be satisfied in this situation.

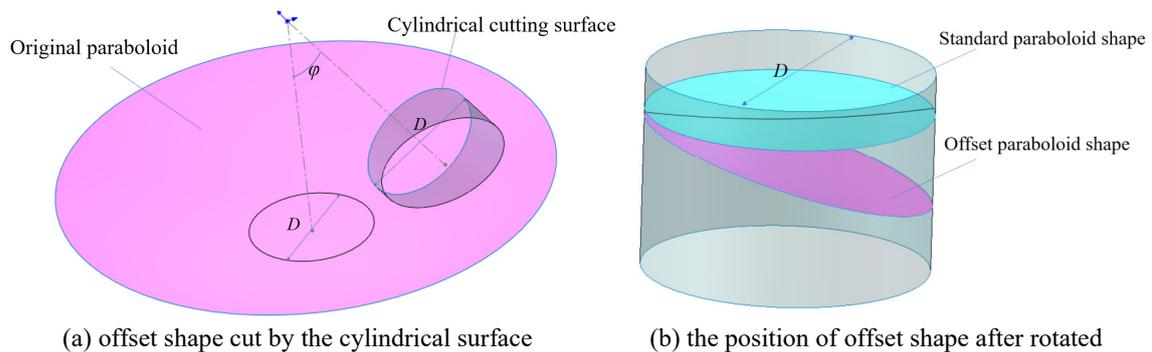

(a) offset shape cut by the cylindrical surface    (b) the position of offset shape after rotated
Fig.2 Offset reflector shape rotated with respect to the focal point

The architecture of the scanning reflector of this study is given here and illustrated in Fig.3:

1) The offset paraboloid shape is obtained by cutting the original paraboloid with an cylindrical cutting surface;

2) The long axis of the offset reflector aperture should be equal to the diameter of the standard reflector aperture, as shown in Fig.3(a), that is, the length of line MN is identical to the length of line AB.

3) Letting the offset shape rotated about the intersecting point (point $O'$) of these two reflector apertures, then the aperture of offset shape is parallel with the standard shape aperture. Here, point S is the middle point of MN and $O'S$ is perpendicular to MN.

4) The offset geometry is then moved alone the Z axis and the points M and N should be coincide with points A and B respectively.

The most advantage of this reconfiguration method is that the deformation becomes much smaller than that of the first method. As a result, it requires an actuator to be much less powerful with a much smaller stoke, as illustrated in Fig.3(c).

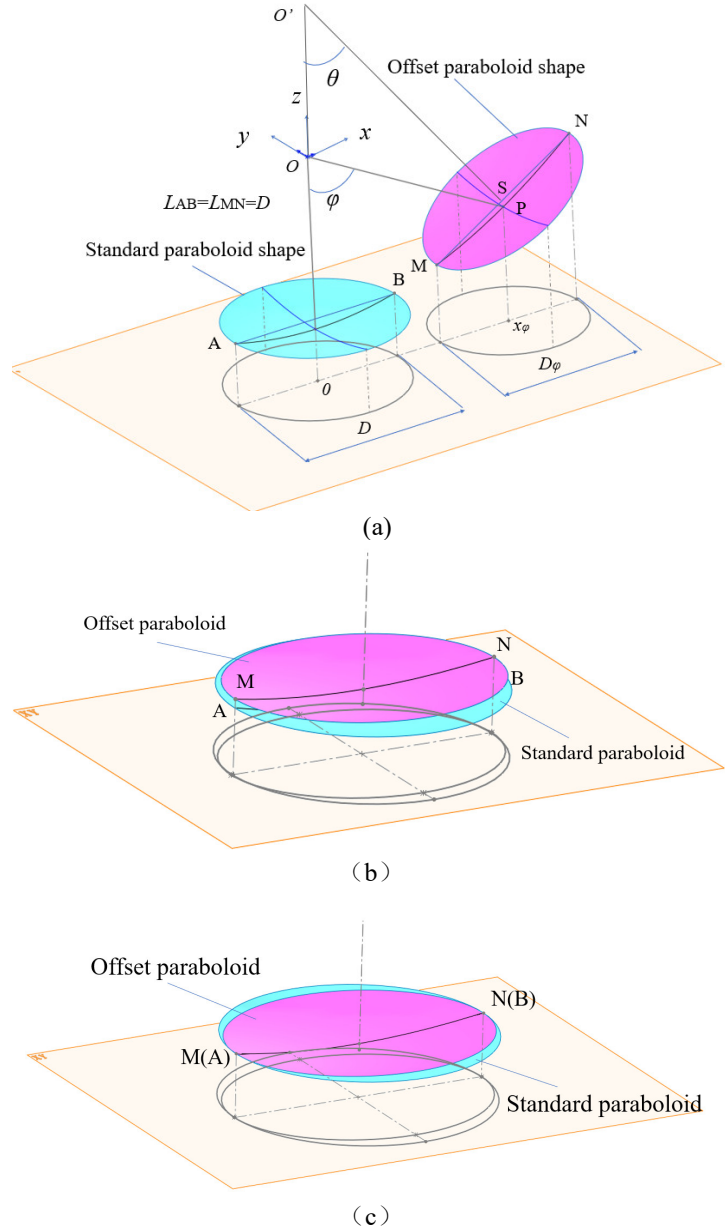

Fig.3 Offset reflector shape rotated about the intersection point

## II. Displacement calculation for the reflector surface reshaping

Assume the original paraboloid equation is:

$$4f(f+z) = x^2 + y^2 \qquad (1)$$

In the polar coordinate system, equation (1) becomes:

$$\rho(\varphi) = f\sec^2\left(\frac{\varphi}{2}\right) \qquad (2)$$

### A. Calculation of the dimeter of cylindrical cutting surface

As shown in Fia.3(a), point P is on the offset paraboloid shape, the corresponding offset angle is $\varphi$, so,

$$x_\varphi = \rho_\varphi \sin\varphi = f\sec^2\frac{\varphi}{2}\sin\varphi = 2f\tan\frac{\varphi}{2} \qquad (3)$$

Assume the diameter of standard reflector aperture is $D$, enable MN=D, so,

$$D_\varphi^2 + (Z_N - Z_M)^2 = D^2 \qquad (4)$$

And,

$$Z_M = f - \frac{\left(x_\varphi - \frac{D_\varphi}{2}\right)^2}{4f}, \; Z_N = f - \frac{\left(x_\varphi + \frac{D_\varphi}{2}\right)^2}{4f} \qquad (5)$$

Combing equation (3), (4) and (5), the dimeter of the cylindrical cutting surface is

$$D_\varphi = \frac{D}{\sqrt{1+\tan^2\frac{\varphi}{2}}} = D\cos\frac{\varphi}{2} \qquad (6)$$

### B. Solving the coordinate value of the rotation center

The coordinate value of the center point of line MN (point S, not same point as P) is

$$(x_\varphi, 0, \frac{Z_M + Z_N}{2}), \text{ i.e. } \left(x_\varphi, 0, f(\tan^2\frac{\varphi}{2}-1) + \frac{D^2\cos^2\frac{\varphi}{2}}{16f}\right)$$

And the slope of line MN is

$$k_1 = \frac{Z_N - Z_M}{D_\varphi} = \tan\frac{\varphi}{2}$$

So, the slope of line MN's midperpendicular is

$$k_2 = -\frac{1}{k_1} = -ctg\frac{\varphi}{2}$$

Obviously, the angle between the midperpendicular and the Z axis is $\theta=\varphi/2$, and the equation of the midperpendicular is

$$z = -xctg\frac{\varphi}{2} + \frac{f}{\cos^2\frac{\varphi}{2}} + \frac{D^2\cos^2\frac{\varphi}{2}}{16f} \qquad (7)$$

And the coordinate value of rotation center $O'$ is

$$\left(0, 0, \frac{f}{\cos^2\frac{\varphi}{2}} + \frac{D^2\cos^2\frac{\varphi}{2}}{16f}\right) \qquad (8)$$

### C. Coordinate transformation of offset shape rotating about point $O'$ to standard shape position

Step1: Origin of coordinates system goes from point O to point O'. Assume the original coordinate of the node on offset shape is $(x_0, y_0, z_0)$, then, it will transform to

$$\begin{pmatrix} x_0 \\ y_0 \\ z_0 \end{pmatrix} - \begin{pmatrix} 0 \\ 0 \\ f/\cos^2\frac{\varphi}{2} + D^2\cos^2\frac{\varphi}{2}/16f \end{pmatrix} = \begin{pmatrix} x_0 \\ y_0 \\ z_0 - f/\cos^2\frac{\varphi}{2} - D^2\cos^2\frac{\varphi}{2}/16f \end{pmatrix} \qquad (9)$$

Likewise, the focal point coordinate will transform to

$$\begin{pmatrix} 0 \\ 0 \\ 0 \end{pmatrix} - \begin{pmatrix} 0 \\ 0 \\ +f/\cos^2\frac{\varphi}{2} + D^2\cos^2\frac{\varphi}{2}/16f \end{pmatrix} = \begin{pmatrix} 0 \\ 0 \\ -f/\cos^2\frac{\varphi}{2} - D^2\cos^2\frac{\varphi}{2}/16f \end{pmatrix} \qquad (10)$$

Step 2: The offset shape rotates about point O' for $\theta=-\varphi/2$. The coordinate of node on offset shape will transform to

$$\begin{bmatrix} \cos\frac{\varphi}{2} & 0 & \sin\frac{\varphi}{2} \\ 0 & 1 & 0 \\ -\sin\frac{\varphi}{2} & 0 & \cos\frac{\varphi}{2} \end{bmatrix} \begin{pmatrix} x_0 \\ y_0 \\ z_0 - \dfrac{f}{\cos^2\frac{\varphi}{2}} - \dfrac{D^2\cos^2\frac{\varphi}{2}}{16f} \end{pmatrix} = \begin{pmatrix} x_0\cos\frac{\varphi}{2} + z_0\sin\frac{\varphi}{2} - f\sin\frac{\varphi}{2}/\cos^2\frac{\varphi}{2} - \dfrac{D^2\sin\frac{\varphi}{2}\cos^2\frac{\varphi}{2}}{16f} \\ y_0 \\ -x_0\sin\frac{\varphi}{2} + z_0\cos\frac{\varphi}{2} - f/\cos\frac{\varphi}{2} - \dfrac{D^2\cos^3\frac{\varphi}{2}}{16f} \end{pmatrix} \quad (11)$$

Likewise, the focal point coordinate will transform to

$$\begin{bmatrix} \cos\frac{\varphi}{2} & 0 & \sin\frac{\varphi}{2} \\ 0 & 1 & 0 \\ -\sin\frac{\varphi}{2} & 0 & \cos\frac{\varphi}{2} \end{bmatrix} \begin{pmatrix} 0 \\ 0 \\ -\dfrac{f}{\cos^2\frac{\varphi}{2}} - \dfrac{D^2\cos^2\frac{\varphi}{2}}{16f} \end{pmatrix} = \begin{pmatrix} -f\sin\frac{\varphi}{2}/\cos^2\frac{\varphi}{2} - \dfrac{D^2\sin\frac{\varphi}{2}\cos^2\frac{\varphi}{2}}{16f} \\ 0 \\ -f/\cos\frac{\varphi}{2} - \dfrac{D^2\cos^3\frac{\varphi}{2}}{16f} \end{pmatrix} \quad (12)$$

In this step, the offset shape is rotated to the position where the apertures of two shape are parallel, as shown in Fig.3(b).

Step 3: Change the origin of coordinates system back to point O. The coordinate of node on offset shape will transform to

$$\begin{pmatrix} x_0\cos\frac{\phi}{2} + z_0\sin\frac{\phi}{2} - f\sin\frac{\phi}{2}/\cos^2\frac{\phi}{2} - \dfrac{D^2\sin\frac{\phi}{2}\cos^2\frac{\phi}{2}}{16f} \\ y_0 \\ -x_0\sin\frac{\phi}{2} + z_0\cos\frac{\phi}{2} - f/\cos\frac{\phi}{2} - \dfrac{D^2\cos^3\frac{\phi}{2}}{16f} \end{pmatrix} + \begin{pmatrix} 0 \\ 0 \\ f/\cos^2\frac{\phi}{2} + \dfrac{D^2\cos^2\frac{\phi}{2}}{16f} \end{pmatrix}$$

$$= \begin{pmatrix} x_0\cos\frac{\phi}{2} + z_0\sin\frac{\phi}{2} - f\sin\frac{\phi}{2}/\cos^2\frac{\phi}{2} - \dfrac{D^2\sin\frac{\phi}{2}\cos^2\frac{\phi}{2}}{16f} \\ y_0 \\ -x_0\sin\frac{\phi}{2} + z_0\cos\frac{\phi}{2} + (1-\cos\frac{\phi}{2})(f/\cos^2\frac{\phi}{2} + \dfrac{D^2\cos^2\frac{\phi}{2}}{16f}) \end{pmatrix} \quad (13)$$

Likewise, the focal point coordinate will transform to

$$\begin{pmatrix} -f\sin\frac{\varphi}{2}/\cos^2\frac{\varphi}{2} - \dfrac{D^2\sin\frac{\varphi}{2}\cos^2\frac{\varphi}{2}}{16f} \\ 0 \\ -f/\cos\frac{\varphi}{2} - \dfrac{D^2\cos^3\frac{\varphi}{2}}{16f} \end{pmatrix} + \begin{pmatrix} 0 \\ 0 \\ \dfrac{f}{\cos^2\frac{\varphi}{2}} + \dfrac{D^2\cos^2\frac{\varphi}{2}}{16f} \end{pmatrix} = \begin{pmatrix} -f\sin\frac{\varphi}{2}/\cos^2\frac{\varphi}{2} - \dfrac{D^2\sin\frac{\varphi}{2}\cos^2\frac{\varphi}{2}}{4f} \\ 0 \\ -f/\cos\frac{\varphi}{2} - \dfrac{D^2\cos^3\frac{\varphi}{2}}{16f} + \dfrac{f}{\cos^2\frac{\varphi}{2}} + \dfrac{D^2\cos^2\frac{\varphi}{2}}{16f} \end{pmatrix} \quad (14)$$

Step 4: Superpose the apertures of offset shape and standard shape. After the coordinate transformation above, according to equation (13) the coordinates of point M and point N transform to

$$\text{M:}\left(-\frac{D}{2},0,-\frac{2f}{\cos\frac{\varphi}{2}}+\frac{f}{\cos^2\frac{\varphi}{2}}+\frac{D^2\cos^2\frac{\varphi}{2}}{16f}\right),\ \text{N:}\left(\frac{D}{2},0,-\frac{2f}{\cos\frac{\varphi}{2}}+\frac{f}{\cos^2\frac{\varphi}{2}}+\frac{D^2\cos^2\frac{\varphi}{2}}{16f}\right)$$

And the coordinate of point A and point B is

$$\text{A:}\left(-\frac{D}{2},0,\frac{D^2}{16f}-f\right),\ \text{B:}\left(\frac{D}{2},0,\frac{D^2}{16f}-f\right)$$

In order to make points M, N superpose with points A, B respectively, the offset shape should move along Z axis for

$$\left(-\frac{2f}{\cos\frac{\varphi}{2}}+\frac{f}{\cos^2\frac{\varphi}{2}}+\frac{D^2\cos^2\frac{\varphi}{2}}{16f}\right)-\left(\frac{D^2}{16f}-f\right)=f(1-\frac{2}{\cos\frac{\varphi}{2}}+\frac{1}{\cos^2\frac{\varphi}{2}})-\frac{D^2\sin^2\frac{\varphi}{2}}{16} \tag{16}$$

And the coordinate of node on offset shape will transform to

$$\begin{pmatrix} x_0\cos\frac{\varphi}{2}+z_0\sin\frac{\varphi}{2}-f\sin\frac{\varphi}{2}/\cos^2\frac{\varphi}{2}-\frac{D^2\sin\frac{\varphi}{2}\cos^2\frac{\varphi}{2}}{16f} \\ y_0 \\ -x_0\sin\frac{\varphi}{2}+z_0\cos\frac{\varphi}{2}+(1-\cos\frac{\varphi}{2})(f/\cos^2\frac{\varphi}{2}+\frac{D^2\cos^2\frac{\varphi}{2}}{16f}) \end{pmatrix} - \begin{pmatrix} 0 \\ 0 \\ f(1-2/\cos\frac{\varphi}{2}+1/\cos^2\frac{\varphi}{2})-\frac{D^2\sin^2\frac{\varphi}{2}}{16f} \end{pmatrix}$$

$$= \begin{pmatrix} x_0\cos\frac{\varphi}{2}+z_0\sin\frac{\varphi}{2}-f\sin\frac{\varphi}{2}/\cos^2\frac{\varphi}{2}-\frac{D^2\sin\frac{\varphi}{2}\cos^2\frac{\varphi}{2}}{16f} \\ y_0 \\ -x_0\sin\frac{\varphi}{2}+z_0\cos\frac{\varphi}{2}-f+\frac{f}{\cos\frac{\varphi}{2}}-\frac{D^2\cos^3\frac{\varphi}{2}}{16f}+\frac{D^2}{16f} \end{pmatrix} \tag{17}$$

The domain of definition of the offset shape is $\left(x_0-x_\varphi\right)^2+y_0^2 \leq D_\varphi^{\ 2}$.

And the focal point coordinate will transform to

$$\begin{pmatrix} -f\sin\frac{\varphi}{2}/\cos^2\frac{\varphi}{2}-\frac{D^2\sin\frac{\varphi}{2}\cos^2\frac{\varphi}{2}}{4f} \\ 0 \\ -f/\cos\frac{\varphi}{2}-\frac{D^2\cos^3\frac{\varphi}{2}}{16f}+\frac{f}{\cos^2\frac{\varphi}{2}}+\frac{D^2\cos^2\frac{\varphi}{2}}{16f} \end{pmatrix} - \begin{pmatrix} 0 \\ 0 \\ f(1-\frac{2}{\cos\frac{\varphi}{2}}+\frac{1}{\cos^2\frac{\varphi}{2}})-\frac{D^2\sin^2\frac{\varphi}{2}}{16f} \end{pmatrix}$$

$$= \begin{pmatrix} -f\sin\frac{\varphi}{2}/\cos^2\frac{\varphi}{2}-\frac{D^2\sin\frac{\varphi}{2}\cos^2\frac{\varphi}{2}}{4f} \\ 0 \\ -f+\frac{f}{\cos\frac{\varphi}{2}}+\frac{D^2}{16f}+\frac{D^2\cos^3\frac{\varphi}{2}}{16f} \end{pmatrix} \tag{18}$$

## D. Displacement calculation of the reconfiguration from standard shape offset shape

Through coordinate transformation above, the node on offset shape $(x_0, y_0, z_0)$ transforms to

$$\begin{pmatrix} x_0 \cos\frac{\phi}{2} + z_0 \sin\frac{\phi}{2} - f\sin\frac{\phi}{2}/\cos^2\frac{\phi}{2} - \frac{D^2 \sin\frac{\phi}{2}\cos^2\frac{\phi}{2}}{16f} \\ y_0 \\ -x_0 \sin\frac{\phi}{2} + z_0 \cos\frac{\phi}{2} - f + \frac{f}{\cos\frac{\phi}{2}} - \frac{D^2 \cos^3\frac{\phi}{2}}{16f} + \frac{D^2}{16f} \end{pmatrix} \quad (19)$$

According to the paraboloid equation (1), the $z$ axis coordinate of the corresponding node on standard shape which has the same coordinate at $x$ and $y$ axis of the node on offset shape is

$$\begin{pmatrix} x_0 \cos\frac{\varphi}{2} + z_0 \sin\frac{\varphi}{2} - f\sin\frac{\varphi}{2}/\cos^2\frac{\varphi}{2} - \frac{D^2 \sin\frac{\varphi}{2}\cos^2\frac{\varphi}{2}}{16f} \\ y_0 \\ -f + \frac{1}{4f}\left((x_0 \cos\frac{\varphi}{2} + z_0 \sin\frac{\varphi}{2} - f\sin\frac{\varphi}{2}/\cos^2\frac{\varphi}{2} - \frac{D^2 \sin\frac{\varphi}{2}\cos^2\frac{\varphi}{2}}{16f})^2 + y_0^2\right) \end{pmatrix} \quad (20)$$

So, the displacement of the node from offset shape to standard shape is

$$dZ = \frac{1}{4f}\left((x_0 \cos\frac{\varphi}{2} + z_0 \sin\frac{\varphi}{2} - f\sin\frac{\varphi}{2}/\cos^2\frac{\varphi}{2} - \frac{D^2 \sin\frac{\varphi}{2}\cos^2\frac{\varphi}{2}}{16f})^2 + y_0^2\right) - \left(-x_0 \sin\frac{\varphi}{2} + z_0 \cos\frac{\varphi}{2} + \frac{f}{\cos\frac{\varphi}{2}} - \frac{D^2 \sin^3\frac{\varphi}{2}}{16f} + \frac{D^2}{16f}\right) \quad (21)$$

## E. Case study

In order to demonstrate the calculation above, a case of a scanning reflector is studied, of which the focus $f$ is 1m, the aperture diameter $D$ is 1m, the scanning range is $\pm 30°$ ($\varphi/2 = 30°$). The original offset reflector shapes are achieved by following equation,

$$\begin{cases} 4f(f+z) = x^2 + y^2 \\ (x - x_{\pi/3})^2 + y^2 \leq D_{\pi/3}^2 \end{cases} \quad (22)$$

Where, $f=1$, and according to equation (3) and (6), $x_{\pi/3} = 2\sqrt{3}/2$, $D_{\pi/3} = \sqrt{3}/2$.

As shown in Fig.4, the original offset reflector shapes are rotated to the standard shape position after the coordinate transformation. In order to compare the standard shape with the transformed offset shape. The standard shape is draw together with the transformed offset shape, and the standard shape is achieved by following equation,

$$\begin{cases} 4f(f+z) = x^2 + y^2 \\ x^2 + y^2 \leq D^2 \end{cases} \quad (23)$$

Where, $D=1$. As shown in Fig.5, the two points in long axis at rim of offset shape are coincide with that of standard shape, which is consistent with the calculation above. It also should be noted that, when the standard shape is reconfigured to the offset shape, some region will be lost.

Also, the displacement at z dimension can be calculated by equation (21), Fig.6 shows the displacement distribution of standard shape's reconfiguration while scanning angle is $\pm 30°$.

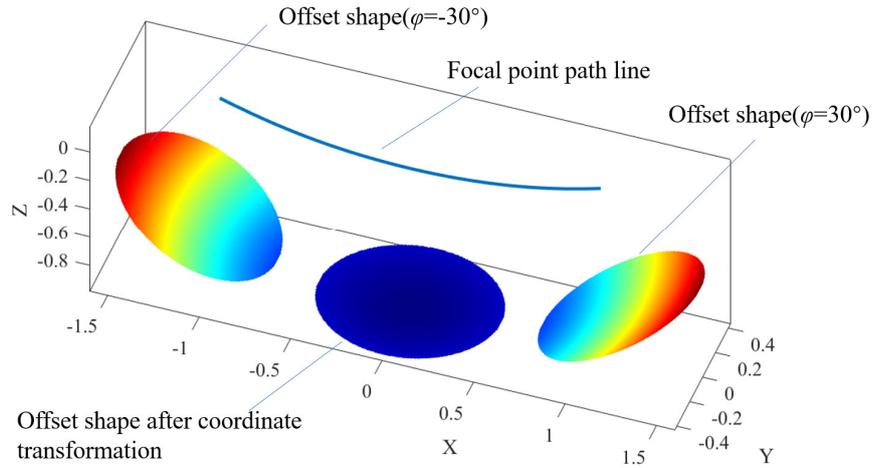

Fig.4 Offset shape before and after coordinate transformation

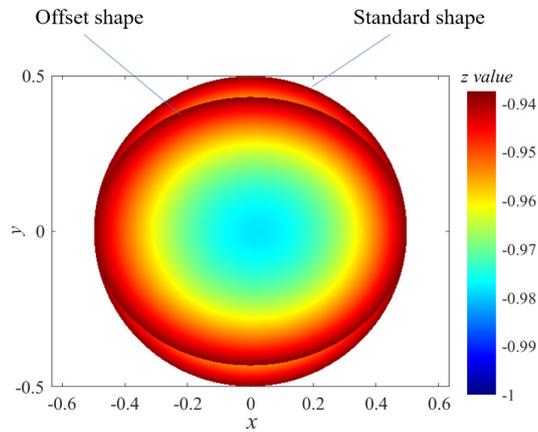

Fig.5 View of standard shape and the transformed offset shape at XY plane

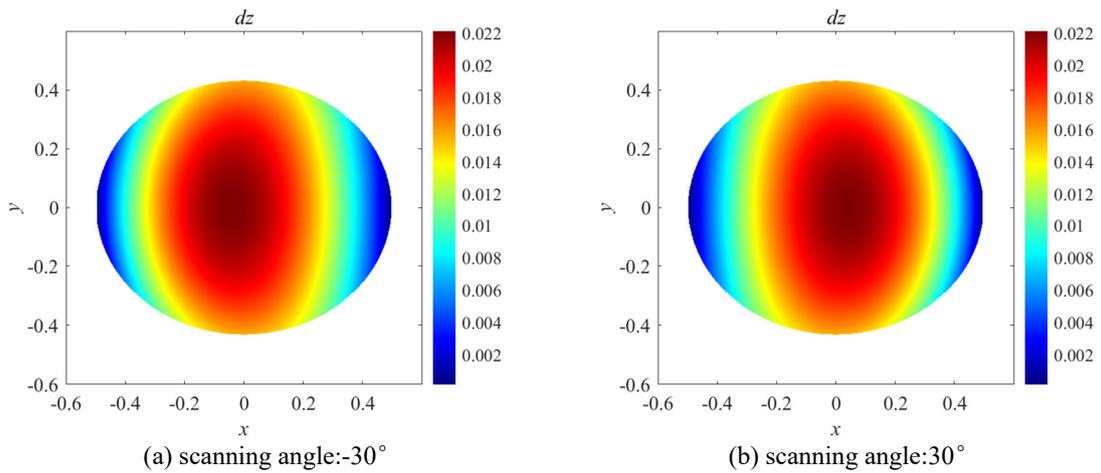

(a) scanning angle:-30°          (b) scanning angle:30°

Fig.6 the displacement of standard shape's reconfiguration

### III. Auxetic material application for mechanical reconfigurable reflector

As the reconfiguration of the scanning antenna reflector is a large elastic out-of-plane deformation, for the uniform material, the actuation force could be large, and results in a large volume and weight of the actuator system. Traditional honey comb structure material is common used as reflector material for its lightweight and low stiffness, which is a good option of mechanical reconfigurable reflector. However, it is also used as a core layer in a sandwich structure in which a top and a bottom skin should be arranged, and wrinkles may easily appear on top or bottom skin cause by compression stress. an NPR lattice with elliptic voids introduced by Bertoldi and co-workers [1] is proposed in this article, see Fig.8.The 2D structure consist of the matrix wherein the elliptical voids are arranged in periodic manner, and can be seen as an analogue of either the rotating square model with square-like elements of the matrix. The elliptical voids have two possible orientations, horizontal and vertical, and the elastic parameters in these two orientations are same [2]. The Poisson's ratio is designable by the adjustment of the geometry parameters $a$, $b$, $g$.

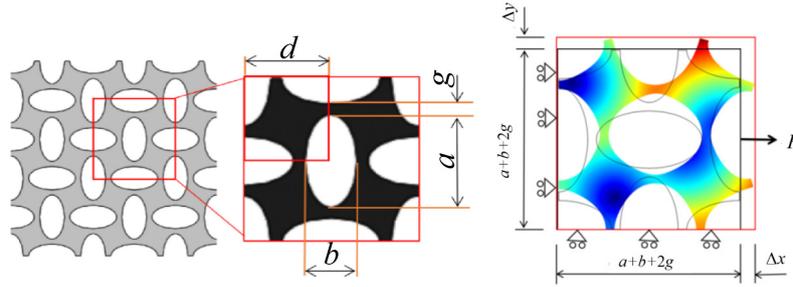

Fig.7 The elliptic voids structure NPR material    Fig.8 FE simulation model

During the optimization, the transmission loss should be taken into consideration, the diameter should be smaller than $\lambda/8$. To lower the tensile stiffness, the parameter g should be small, but that will increase the manufacturing difficulty and cost. A FE model, as shown in Fig.8, is used to simulate the deformation and the effective Poisson's ratio $v_{eff}$ and the ratio of effective Young's modulus ($E_{eff}$) to the Young's modulus of material ($E_{mat}$) are calculated by the following way:

$$v_{eff} = -\frac{\varepsilon_y}{\varepsilon_x} = -\frac{\Delta y/(a+b+2g)}{\Delta x/(a+b+2g)} = -\frac{\Delta y}{\Delta x} \qquad (24)$$

$$\frac{E_{eff}}{E_{mat}} = \frac{\sigma/\varepsilon_x}{E_{mat}} = \frac{\dfrac{F}{(a+b+2g)t} \dfrac{a+b+2g}{\Delta x}}{E_{mat}} = \frac{F/(t\Delta x)}{E_{mat}} \qquad (25)$$

The material investigated is carbon fibre reinforced silicon (CFRS), for example, Young's modulus $E_{mat}$=500MPa, Poisson's ratio is 0.4. The simulation results are shown in Fig.9. With the ratio of $b/a$ increasing, the ratio of $E_{eff}/E_{mat}$ and $v_{eff}$ increase.

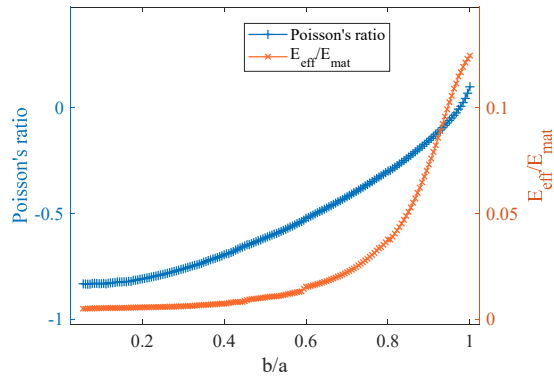

Fig.9 Effective Poisson's ratio vs. $b/a$

In order to find the relationship of the ratio of b/a with the out-of-plane deformation, the simulation of deformation of reflector with structure of NPR lattice of elliptic voids under the action of point force is taken. The point force of 10N is applied at the center of the reflector, and the rim is fixed, see Fig.10(a). It can be seen that the displacement increases with the ratio of b/a, in other words, the bending stiffness decrease.

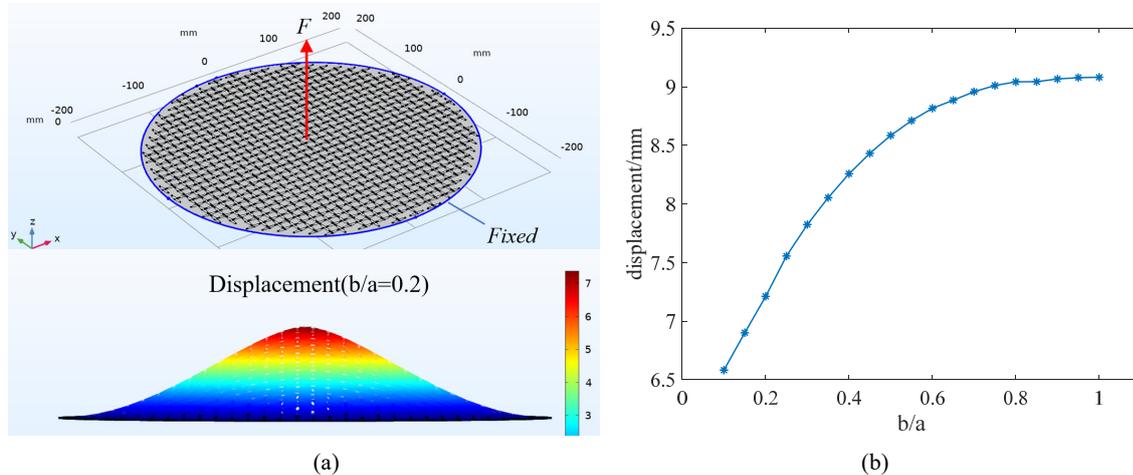

(a)          (b)

Fig.10 Simulation of deformation of reflector with structure of NPR lattice of elliptic voids under the action of point force (a) boundary setting and result when b/a=0.2 (b) the maximum displacement *vs*. b/a

According to the simulations above, while increase *b/a* of the elliptic voids, the effective Young's modulus increase, and the bending stiffness decrease. Thus, the ratio of bending stiffness to membrane stiffness decrease. According to the research of Leri Datashvili [3,4], for the material properties of a flexible reflecting morphing skin, the ratio of bending stiffness to membrane stiffness must be high to achieve smooth reshaping. So, the ratio of *b/a* should be as small as possible.

## IV. Conclusion

The proposed reconfiguration method for the scanning antenna reflector can save much stroke for the actuator during surface reshaping. The coordinate of the offset paraboloid which the standard paraboloid reshape to is calculated by coordinate transformation and the displacement of actuator can be achieved. A NPR lattice structure with elliptic voids applied to the reconfigurable reflector can decrease membrane stiffness and make bending stiffness dominated in the material property, that can achieve smooth reshaping.